\pdfoutput=1

\documentclass[11pt]{article}

\usepackage[final]{acl}

\usepackage{times}
\usepackage{latexsym}

\usepackage[T1]{fontenc}

\usepackage[utf8]{inputenc}

\usepackage{microtype}

\usepackage{inconsolata}

\usepackage{graphicx}
\usepackage{algorithm2e}
\usepackage{amsmath}
\usepackage{booktabs}
\usepackage{multirow}
\usepackage{lipsum}
\usepackage{pifont}
\usepackage{makecell}
\usepackage{tabularx}
\usepackage{subcaption}
\usepackage{enumitem} 

\usepackage{amsfonts} 
\usepackage[normalem]{ulem}
\usepackage{bm}

\usepackage[table]{xcolor}
\usepackage{colortbl}
\usepackage{pgf}
\usepackage{xfp} 
\usepackage{collcell}   
\usepackage{colortbl}   
\usepackage{xcolor}     
\usepackage{pgfmath}    
\usepackage{array}
\usepackage{tabularx}

\definecolor{biasLow}{RGB}{255,255,255}   
\definecolor{biasHigh}{RGB}{255,130,60}   

\title{Stop Playing the Guessing Game! Target-free User Simulation for Evaluating Conversational Recommender Systems}

\author{
    Sunghwan Kim\thanks{\; Equal contribution}~~~
    Kwangwook Seo$^\ast$~~~
    Tongyoung Kim$^\ast$~~~
    Jinyoung Yeo~~~
    Dongha Lee\thanks{\; Corresponding author}\\
    Department of Artificial Intelligence, Yonsei University\\
    \texttt{\{happysnail06,tommy2130,dykim,jinyeo,donalee\}@yonsei.ac.kr}\\   
}
\newcommand{\redial}{Redial\xspace}
\newcommand{\opendialkg}{OpenDialKG\xspace}
\newcommand{\imdb}{IMDB\xspace}

\newcommand{\proposed}{PEPPER\xspace}

\newcommand{\pc}{\textsc{Preference Coverage}\xspace}
\newcommand{\pcir}{\textsc{Preference Coverage Increase Rate}\xspace}

\newcommand{\proact}{\textit{Proactiveness}\xspace}
\newcommand{\coh}{\textit{Coherence}\xspace}
\newcommand{\personal}{\textit{Personalization}\xspace}

\newcommand{\kbrd}{KBRD\xspace}
\newcommand{\barcor}{BARCOR\xspace}
\newcommand{\unicrs}{UniCRS\xspace}
\newcommand{\chatgpt}{ChatGPT\xspace}

\newcommand{\chatcrs}{ChatCRS\xspace}
\newcommand{\macrs}{MACRS\xspace}

\newcommand{\ievalm}{iEvaLM\xspace}
\newcommand{\simusersim}{SimpleUserSim\xspace}
\newcommand{\cshi}{CSHI\xspace}
\newcommand{\concept}{CONCEPT\xspace}

\definecolor{DarkGreen}{RGB}{30,130,30}
\newcommand{\cmark}{\textcolor{DarkGreen}{\ding{51}}}
\newcommand{\xmark}{\textcolor{red}{\ding{55}}}%

\begin{document}
\maketitle

\begin{abstract}
Recent developments in Conversational Recommender Systems (CRSs) have focused on simulating real-world interactions between users and CRSs to create more realistic evaluation environments. 
Despite considerable advancements, reliably assessing the capability of CRSs in eliciting user preferences remains a significant challenge. 
We observe that user-CRS interactions in existing evaluation protocols resemble a \textit{\textbf{guessing game}}, as they construct \textit{target-biased }simulators pre-encoded with target item knowledge, thereby allowing the CRS to shortcut the elicitation process.
Moreover, we reveal that current evaluation metrics, which predominantly emphasize single-turn recall of target items, suffer from \textit{\textbf{target ambiguity}} in multi-turn settings and overlook the intermediate process of preference elicitation.
To address these issues, we introduce \textbf{{\proposed}}, a novel CRS evaluation protocol with \textit{target-free} user simulators that enable users to gradually discover their preferences through enriched interactions, along with detailed measures for comprehensively assessing the preference elicitation capabilities of CRSs.
Through extensive experiments, we validate \proposed as a reliable evaluation protocol and offer a thorough analysis of how effectively current CRSs perform in preference elicitation and recommendation.
\url{https://github.com/happysnail06/PEPPER}
\end{abstract}

\section{Introduction}
\label{sec:intro}
\begin{figure}[!t]
    \centering
    \includegraphics[width=1\columnwidth]{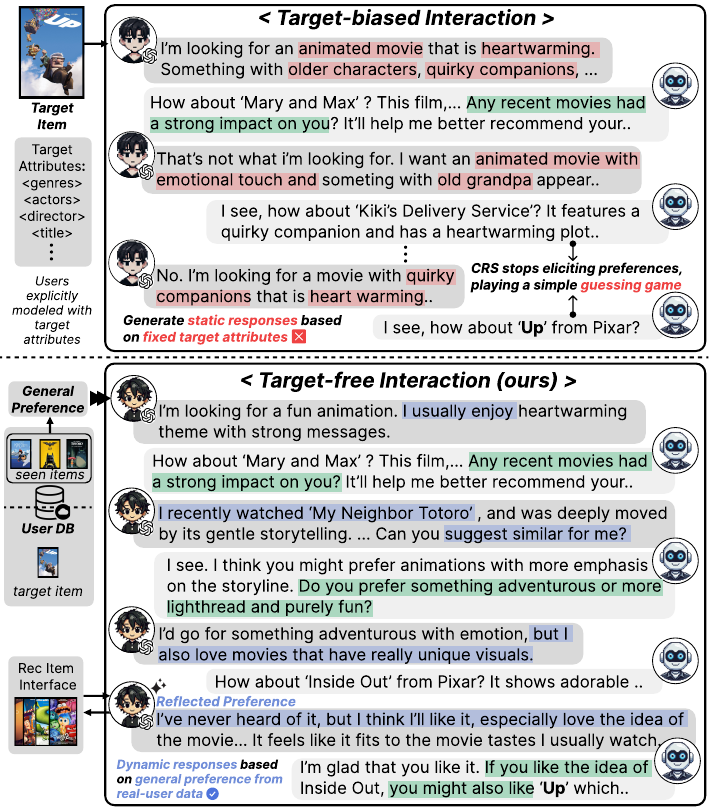}
    \caption{While existing \textit{target-biased} user simulators directly reveal attributes of target items for CRS evaluation (Upper), our \textit{target-free} user simulator engages with more general preference (Lower), making preference elicitation crucial to provide accurate recommendations.
    }
    \vspace{-0.3cm}
    \label{fig:motivating} 
    \vspace{-0.35cm}
\end{figure} 
Conversational recommender systems (CRSs) have played an increasingly important role in enhancing personalized experiences by providing tailored recommendations through interactive dialogues~\citep{sun2018conversational,jannach2021survey,lin2023enhancing}. 
Throughout the interaction, these systems are expected to perform two key tasks: (1) \textit{preference elicitation} - exploring and uncovering user preferences by encouraging them to express their likes and dislikes, and (2) \textit{recommendation} - retrieving personalized items based on the preferences inferred from the dialogue.
%
In the field of CRSs, automatically evaluating the system's capability has remained challenging~\citep{friedman2023leveraging,wu2024survey,zhao2024recommender,lin2023can,zhu2024reliable}.
Conventional offline approaches relying on static, pre-collected dialogues from datasets often neglect the system’s responsibility to dynamically shape the dialogue itself, whereas evaluating with real user interactions is costly and time-consuming~\citep{zhang2020evaluating,Gao_2021,yoon2024evaluating}.

Recently, many studies~\citep{zhang2020evaluating, friedman2023leveraging} have explored leveraging Large Language Models (LLMs) to simulate user conversations with CRSs, creating more realistic evaluation environments that reflect the complexity of human-agent dialogue.
However, while effective at assessing recommendation quality, these approaches still face challenges in reliably evaluating the process of preference elicitation.
Specifically, we highlight two major limitations in existing user simulation paradigms:
\textbf{(1) Target-biased user simulation}: 
Existing methods assume scenarios where users have specific items in mind, thereby constructing user simulators that are explicitly informed by the target item attributes.
However, relying on the target items to model the user simulator significantly hinders user-CRS interactions, as it tends to generate static responses that repeatedly expose the same target attributes, causing the CRS to take shortcuts to the target items.
\textbf{ (2) Lack of reliable metrics}: 
Existing evaluation metrics are typically limited to measuring single-turn recall of target items, without accounting for the intermediate elicitation process.
As a result, they fail to fully assess how well the CRS guides the conversation to uncover the user's evolving preferences or how effectively it addresses the user's diverse tastes throughout the interaction.

Motivated by these, this paper begins by investigating two key research questions: \textbf{(1) How does reliance on target items affect the quality of user-CRS interactions?}
We reveal that target-biased user simulators reduce interactions to a simplistic {\textit{\textbf{guessing game}}}~\citep{yoon2024evaluating}, where the CRS succeeds by repeatedly guessing the target items rather than meaningfully eliciting user preferences. This oversimplified interaction inflates CRS performance and leads to substantial performance disparities across target items, ultimately distorting evaluation results.
(Figure~\ref{fig:motivating} Upper).
\textbf{(2) How reliable is $\text{Recall}@K$ as a metric for evaluating CRS in multi-turn dialogues?} 
We observe that $\text{Recall}@K$ suffers from \textit{\textbf{target ambiguity}} in multi-turn settings, where the system may hit different target items at each turn yet receive the same score—failing to capture meaningful differences in recommendation behavior.
This limitation makes it difficult to distinguish whether the CRS is genuinely guiding the conversation to uncover new target items or merely reiterating previous recommendations.

To tackle these challenges, we propose a novel \underline{\textbf{P}}rotocol for \underline{\textbf{E}}valuating \underline{\textbf{P}}ersonal \underline{\textbf{P}}reference \underline{\textbf{E}}licitation and \underline{\textbf{R}}ecommendation of CRS,
named \textbf{\proposed}.
To address the target-biased interactions of user simulators, \proposed adopts \textit{target-free} user simulators, modeled on diverse preferences drawn from real user interaction histories and reviews.
Built upon real user data, our simulators personalize their initial behavior based on the review-driven user profiles, instead of relying on fixed target item attributes. 
In particular, we encourage users to actively participate in conversations with the CRS, enabling them to gradually discover their own preferences through interaction (Figure~\ref{fig:motivating} Lower).
To achieve this, we simulate users to continuously enrich the responses by incorporating implicit preferences derived from reflecting their general preferences on items emerging within the interaction.

Moreover, we introduce both quantitative and qualitative measures to comprehensively evaluate preference elicitation capabilities of CRSs.
For quantitative measure, we propose a new metric, \pc, to assess how effectively the CRS elicits each user's diverse preferences with high coverage evolving throughout the conversation.
For qualitative measure, we propose fine-grained scoring rubrics to evaluate three different aspects of preference elicitation: \textit{proactiveness}, \textit{coherence} and \textit{personalization}. 

To summarize, our contributions are as follows: 
\begin{itemize}[leftmargin=23pt, nosep]
    \item We provide detailed analysis of two key limitations in existing CRS evaluation protocols: (1) target-biased user simulation and (2) lack of reliable metrics.
    
    \item We propose {\proposed}, a novel CRS evaluation protocol with \textit{target-free} user simulators, enabling realistic user-CRS dialogues without falling into simplistic \textit{guessing games}.

    \item We present detailed measures for comprehensively evaluating the preference elicitation capabilities of CRSs, encompassing both quantitative and qualitative approaches.
    
    \item Through extensive experiments, we demonstrate the validity of \proposed as a simulation environment and conduct a thorough analysis of how effectively existing CRSs perform in preference elicitation and recommendation. 
\end{itemize}

\begin{table*}[t]
\centering
\resizebox{0.99\linewidth}{!}
{
\begin{tabular}{cccccccc}
    \toprule
    \multirow{2.5}{*}{\textbf{Method}} & \multirow{2.5}{*}{\makecell{\textbf{Dataset}\\\textbf{(Movie Domain)}}} & \multicolumn{4}{c}{\textbf{User Simulation}} & \multicolumn{2}{c}{\textbf{CRS Evaluation}} \\ 
    \cmidrule(lr){3-6} \cmidrule(lr){7-8}
    & & \multicolumn{1}{c}{\textbf{User Profile Input}} & \textbf{Target-free} & \textbf{Free-form} & \textbf{Interaction Strategy} & \textbf{Pref. Elicit.} & \textbf{Recommend.} \\
    \midrule
    \ievalm~\cite{wang2023rethinking} & \redial, \opendialkg & Target Item Title & \xmark & \xmark & \xmark & \xmark & \cmark \\
    \simusersim~\cite{zhu2024reliable} & \redial, \opendialkg & Target Item Attr. & \xmark & \xmark & \xmark & \xmark & \cmark \\
    \cshi~\cite{zhu2025llm} & MovieLens & Target Item Attr., Long-term Pref. & \xmark & \cmark & Intent Understanding  & \xmark & \cmark \\
    \concept~\cite{huang2024concept} & LLM-Generated & Target Item Attr., Personality & \xmark & \cmark & Feeling Generation & \xmark & \cmark \\
    \midrule
    \proposed (Ours) & IMDB & General Preference & \cmark & \cmark & Preference Reflection & \cmark & \cmark \\
    \bottomrule
    \end{tabular}
}
\vspace{-0.2cm}
\caption{
Comparison of existing CRS evaluation protocols with LLM-based user simulators. }
\vspace{-0.3cm}
\label{tbl:comparison}
\end{table*}

\section{Related Work}
\label{sec:relatedwork}
\subsection{Conversational Recommender Systems}
Conversational Recommender System (CRS) aims to elicit user preferences and provide personalized recommendations through conversations. 
In the field of CRSs, one line of research ~\cite{wang2022barcor, wang2022towards} has focused on refining architectural designs to improve recommendation accuracy, while another ~\cite{kostric2021soliciting, ziegfeld2025effect} has emphasized enhancing the preference elicitation process to support more personalized interactions.
Despite significant advancements, previous evaluation protocols have predominantly focused on measuring final recommendation accuracy using pre-collected dialogue datasets ~\cite{chen2019towards, wang2022towards, wang2022barcor}, often overlooking the interactive process of preference elicitation.
Consequently, automatic evaluation of CRSs has emerged as a key challenge in CRS, as it requires to create more realistic testing environments that reflect the complexity of human-agent dialogue. 
\subsection{CRS Evaluation with User Simulator}

Recently, researchers have focused on developing user simulators for evaluating the performance of CRSs~\cite{zhang2020evaluating, yoon2024evaluating}.
iEvaLM~\cite{wang2023rethinking} addresses the limitations of traditional offline evaluation methods by dynamically extending pre-collected dialogues through free-form interactions.
While effective, concerns have been raised about data leakage, where target item titles are disclosed in existing dialogue histories or user prompt, leading to inflated evaluation results.
To mitigate this, ~\cite{zhu2024reliable, huang2024concept, zhu2025llm} have tried to model user preferences using only target item attributes (e.g., genres).
However, this simplification still falls short of fully addressing the core issue, as providing target attributes can still shortcut the recommendation process by implicitly narrowing the candidate space.
A summary of the existing simulation methods is shown in Table~\ref{tbl:comparison}.
\section{Preliminary Analysis}
\subsection{\textbf{Focus and Task}}
\medskip
\noindent\textbf{Focus:}
We focus on unveiling the impact of target-biased user simulation and the limitations of current evaluation metrics in assessing CRS performance. Specifically, we analyze how (1) reliance on predefined target items and (2) the use of Recall as an evaluation metric distort the evaluation process.

\medskip
\noindent\textbf{Task: } CRSs aim to identify a user’s target items through multi-turn, preference-eliciting dialogues.
Formally, given a user-item dataset, $\mathcal{U}$ and $\mathcal{I}$ denote the sets of users and items, respectively. For each user $u \in \mathcal{U}$, the preference is modeled with a set of target items $i_u \subset \mathcal{I}$.
During interaction, the user provides utterances $u_t$ at each turn, either stating their preferences or giving feedback on prior recommendations. 
The CRS then generates a response $r_t$ along with a predicted item list $P_t \subset \mathcal{I}$.
The ultimate goal of the CRS is to recommend items contained in the user's target set $i_u$.

\subsection{Evaluation Setup}
\noindent\textbf{Dataset. }We use \imdb\footnote{https://www.imdb.com/} movie dataset to initialize user simulators and conduct our experiments on CRSs trained with \redial~\cite{li2018towards} and \opendialkg~\cite{moon2019opendialkg} datasets.
To ensure a reliable evaluation, we have aligned movie entities in \imdb with each CRS dataset by retaining only the items shared between them.
Further details on the dataset is described in \ref{appendix:dataset}.

\medskip
\noindent\textbf{Metric.} To reflect how the CRS performs throughout the interaction, we use $\text{Recall}@(t, K)$, which measures the proportion of target items successfully retrieved at the $t$-th turn.

\medskip
\noindent\textbf{CRS Baselines.}
We evaluate six representative CRSs, including three supervised models—\kbrd, \barcor, and \unicrs—and three LLM-based methods.
The implementation details of these models are provided in Appendix \ref{appendix:crs}

\medskip
\noindent\textbf{Target-biased User Simulation.}
Following \cite{zhu2024reliable}, we initialize the preferences of the target-biased user simulators by excluding movie titles and relying solely on item attribute information (i.e., genres, directors, stars, and plot summaries). 
To explore how target-item reliance impacts user-CRS interaction, we further divide the target item set into two parts: a randomly sampled subset, denoted as the \textit{selected} set, and the remaining subset, denoted as the \textit{residual} set.
We then implement target-biased user simulators using only the attributes from the \textit{selected} set.
We hypothesize that user preferences modeled solely from the selected target attributes fail to fully capture the diversity of human interests. 
Otherwise, such attribute-based representations would be sufficient to generalize and allow the CRS to discover the full range of target items, including the \textit{residual} set.
To examine this, we compare CRS performance on the \textit{selected} and \textit{residual} sets.
Further implementation details are provided in Appendix~\ref{appendix:biasuser}.

\subsection{\textbf{Comparison of Residual \& Selected sets}}
To verify that \textit{residual }and \textit{selected} sets are fairly split, we investigate attribute-level similarity between the two sets. 
Specifically, we analyze two complementary aspects:
(1) Measuring categorical overlap of attribute (i.e., genre) by computing the jaccard similarity between \textit{residual} and \textit{selected} items.
(2) Comparing semantic similarity of attribute (i.e., plot) between the residual and selected items.
To achieve this, we compare four pair types: (i) Intra-Genre (upper-bound similarity within a genre bucket), (ii) Inter-Genre (lower-bound similarity across genres), (iii) Seen–Seen (a random half-split of the user’s watched movies), and (iv) our Selected-Residual pairs.
As shown in Table~\ref{tbl:sel_res_similarity}, in both datasets and metrics, Selected–Residual similarity almost exactly matches Seen–Seen similarity. 
This indicates that splitting is not skewed toward specific item attribute patterns.
In both dimensions, Selected–Residual similarity falls comfortably between the lower and upper bounds (Inter-Genre < Selected–Residual < Intra-Genre), avoiding both excessive similarity that could trivialize the task and excessive dissimilarity that could compromise evaluation fairness.
\setlength{\tabcolsep}{2pt}
\renewcommand{\arraystretch}{1.1}
\newcolumntype{C}{@{\hspace{3pt}}c@{\hspace{3pt}}}

\begin{table}[h]
\centering
\footnotesize
\resizebox{\columnwidth}{!}{
\begin{tabular}{l l CCC}
\toprule
\textbf{Dataset} & \textbf{Comparison} &
\textbf{Genre Sim.} & \textbf{Plot Sim.} & \textbf{Combined Sim.} \\
\midrule
\multirow{4}{*}{\rotatebox{90}{\parbox[c][1cm][c]{1cm}{\centering IMDB \\ \textsuperscript{ReDial}}}}
  & Intra-Genre         & 0.3557 & 0.2461 & 0.3119 \\
  & Inter-Genre         & 0.1406 & 0.2037 & 0.1658 \\
  & Seen-Seen           & 0.2245 & 0.2378 & 0.2298 \\
  & Selected-Residual   & 0.2220 & 0.2190 & 0.2210 \\
\midrule
\multirow{4}{*}{\rotatebox{90}{\parbox[c][1cm][c]{1cm}{\centering IMDB \\ \textsuperscript{OpenDialKG}}}}
  & Intra-Genre         & 0.3598 & 0.2337 & 0.3093 \\
  & Inter-Genre         & 0.1523 & 0.1958 & 0.1697 \\
  & Seen–Seen           & 0.2307 & 0.2286 & 0.2299 \\
  & Selected-Residual   & 0.2231 & 0.2327 & 0.2269 \\
\bottomrule
\end{tabular}}
\vspace{-0.3cm}
\caption{Mean similarity scores across baselines and critical comparisons.
\textbf{Intra-Genre}: similarity between movie pairs within the same genre group.
\textbf{Inter-Genre}: similarity between movie pairs from different genre groups.
\textbf{Seen-Seen}: similarity between two halves of each user's watched movies (internal consistency).
\textbf{Selected-Residual}: similarity between selected and non-selected movies in recommendation experiments.}
\vspace{-0.2cm}
\label{tbl:sel_res_similarity}
\end{table}

\begin{figure}[!t]
    \centering
    \includegraphics[width=1\columnwidth]{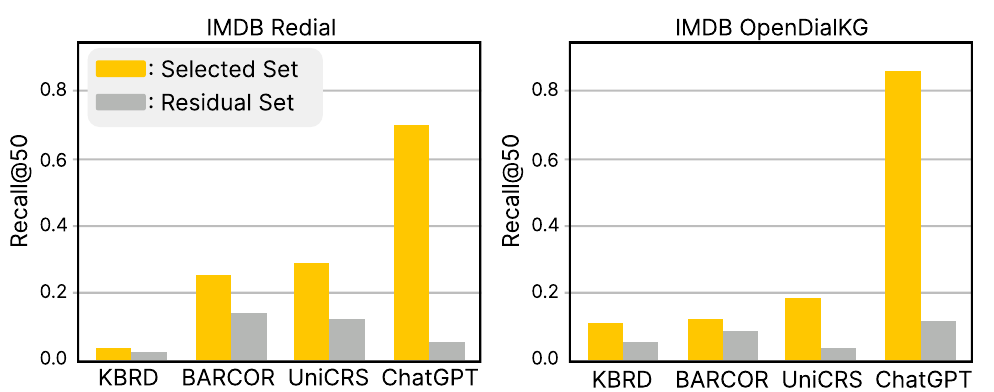}
    \vspace{-0.3cm}
    \caption{Comparison between selected and residual recall for revealing target-item reliance in user simulators.
    }
    \vspace{-0.3cm}
    \label{fig:residual_select_recall}  
\end{figure} 
\subsection{\textbf{Results and Analysis}}
\label{sec:analysis}
\textbf{Target-biased user simulation results in a guessing game}. As shown in Figure~\ref{fig:residual_select_recall}, the results reveal a significant performance disparity for target-biased user simulation.
For example, on the IMDB\textsubscript{OpenDialKG} dataset, ChatGPT achieves an average score of 0.86 for the \textit{selected} set but only 0.12 for the \textit{residual} set.
Similar trends are observed in other CRS models and in the results from the IMDB\textsubscript{ReDial} dataset, further confirming the presence of significant bias.
We interpret this bias as a consequence of target disclosure, where target-biased user simulators tend to prioritize certain target items based on their known attributes, resulting in static and narrowly focused preferences that fail to generalize to the \textit{residual} set.
Moreover, target-biased simulators tend to provide shortcuts for CRSs by explicitly revealing the target item attributes, reducing the need for meaningful preference elicitation and substantially inflating evaluation results.
This calls into question the reliability of existing evaluation protocols and highlights the need for a more realistic user simulation approach.

\begin{figure}[h]
    \centering
    \includegraphics[width=1\columnwidth]{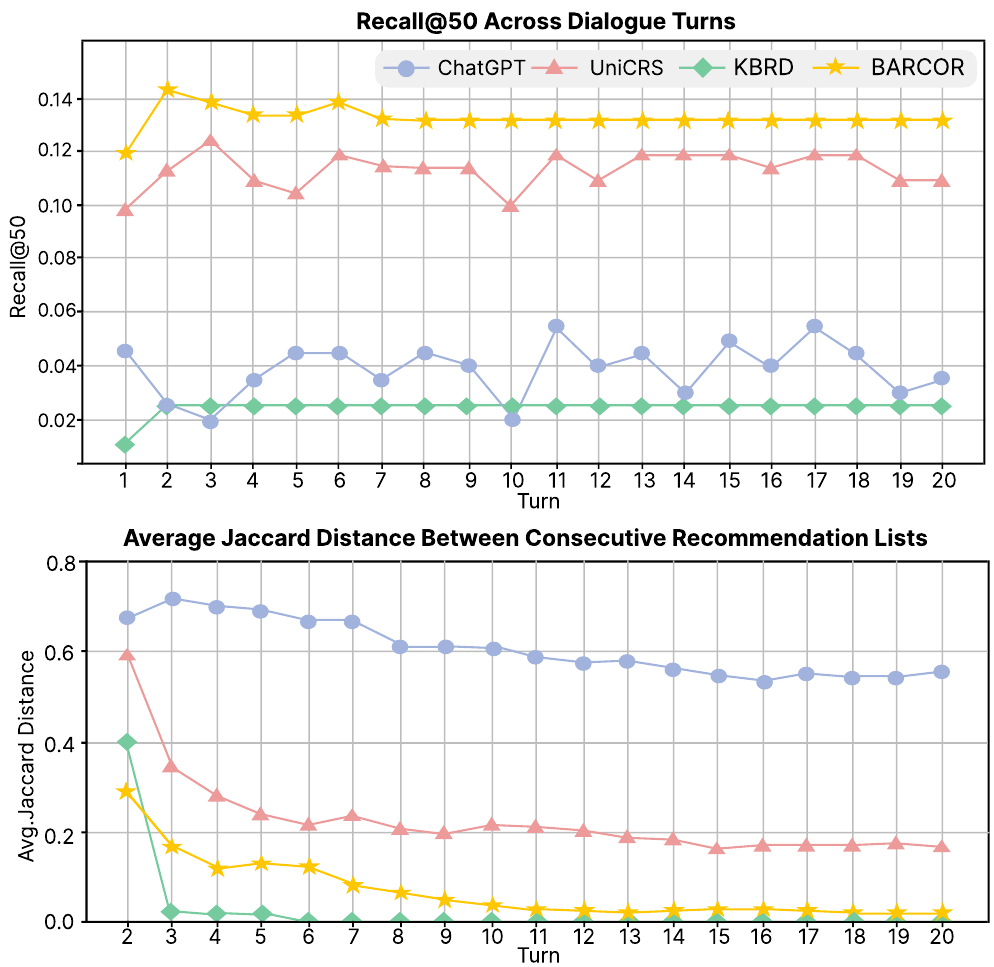}
    \vspace{-0.3cm}
    \caption{(Upper) Recall@50 of the different CRSs across 20 dialogue turns on the IMDB\textsubscript{ReDial} dataset. (Lower) Average Jaccard distance between consecutive recommendation lists of CRS at each turn.}
    \vspace{-0.3cm}
    \label{fig:CLR@50} 
\end{figure}

\medskip
\paragraph{Recall@$K$ fails to reflect meaningful preference elicitation.}
Preference elicitation in conversational recommendation involves progressively uncovering users' diverse preferences through interactive dialogue. 
However, relying solely on Recall exhibits structural limitations that prevent it from properly reflecting this elicitation process.
Specifically, Recall@$K$ (1) permits redundancy by allowing repeated counting of identical items across turns (refer to as \textit{target ambiguity}) and (2) measures performance independently at each turn, ignoring previously discovered or missed preferences.
For example, as shown in Figure~\ref{fig:CLR@50}, ChatGPT consistently explores new items at each turn, indicated by its high Jaccard distance, whereas KBRD rarely updates its recommendations (low Jaccard distance). 
Although ChatGPT actively explores new preferences, Recall@$K$ captures only the low hit rate per turn, failing to acknowledge its consistent efforts and  treating both models similarly, despite substantial differences in their preference exploration behaviors.
Therefore, Recall@$K$ alone fails to capture the process of preference elicitation and points to the need for a metric that reflects diverse preference discovery throughout the dialogue. 
\label{sec:recall_fails}

\section{PEPPER: Target-free CRS Evaluation}
\label{sec:method}
\begin{figure*}[!ht]
    \centering
    \includegraphics[width=1\textwidth]{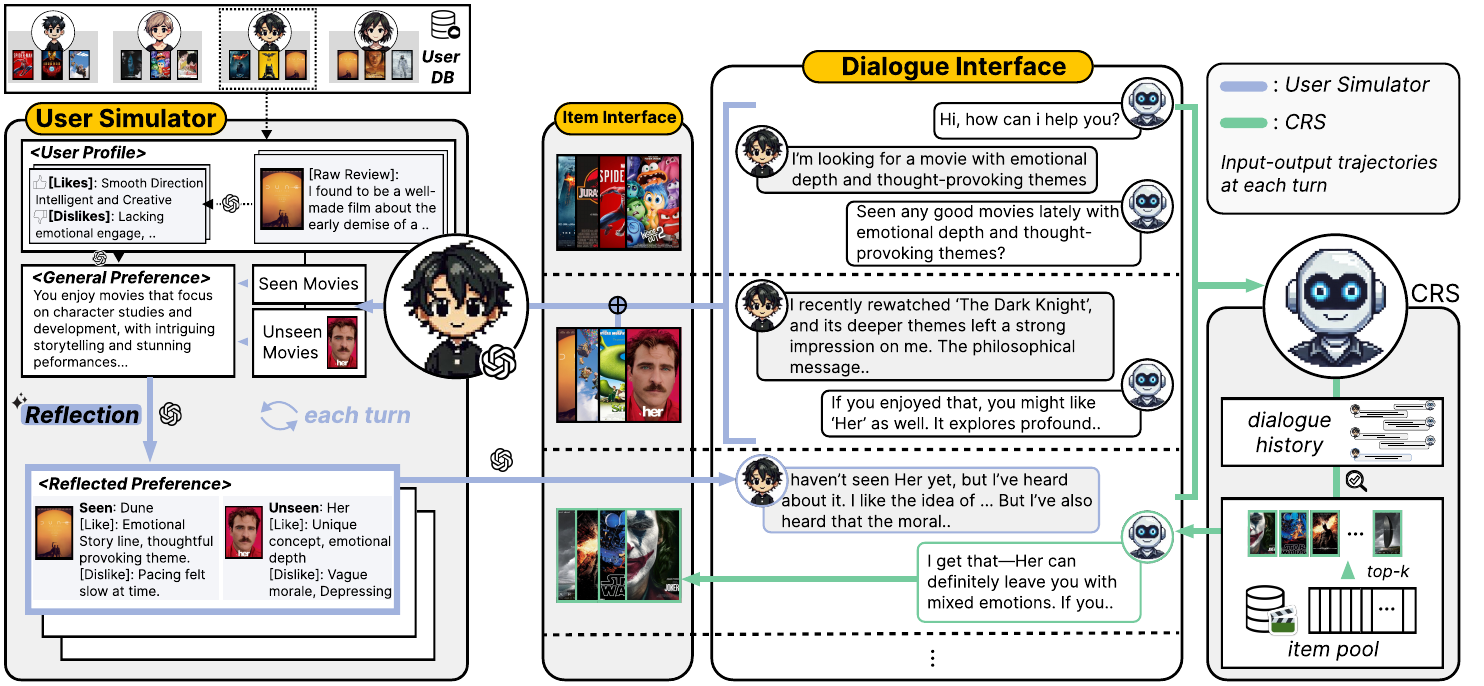}
    \caption{Overview of \proposed. Within our protocol, a user simulator and a CRS interact via (1) item interface and (2) dialogue interface. The user simulator is initialized with general preferences derived from real-world datasets (\textit{i.e.}, IMDB). [\textit{Blue line}] At each interaction, the user simulator first inspects top-$k$ recommendations in the item interface, classifying the items into seen and unseen sets. It then uses these classifications and the general preferences to generate reflected preferences. Finally, it provides a tailored response enriched with detailed personal preferences. [\textit{Green line}] In response, the CRS generates an utterance and presents new item recommendations.
    }
    \vspace{-0.2cm}
    \label{fig:overview}
    \vspace{-0.3cm}
\end{figure*}

Guided by the limitations of existing evaluation protocols, we introduce \proposed, a novel evaluation protocol designed to comprehensively assess both preference elicitation and recommendation abilities of CRSs, addressing key shortcomings of prior approaches.
Specifically, it incorporates two key components: (1) \textbf{\textit{target-free user simulators}} with richly expressed preferences derived from real user interaction histories and reviews, and (2) \textbf{\textit{preference elicitation metrics}} that thoroughly measure a CRS's ability to uncover diverse user preferences and deliver accurate recommendations.

\subsection{Target-free User Simulator} \label{sec:targetfree}
Unlike prior approaches \cite{wang2023rethinking, zhu2024reliable, zhu2025llm, huang2024concept}, which assume scenarios where users have predefined target items in mind, we design our user simulators with diverse preferences derived from actual user experiences.
We aim to construct target-free simulators, instructing them to seek target items without any predefined target information. Instead, these user simulators gradually elaborate on their preferences through ongoing conversations, mirroring how real users naturally articulate and discover their interests. 
To achieve this, we introduce two core components: \textbf{\textit{General Preferences}} and \textbf{\textit{Reflected Preferences}}.
Specifically, general preferences are established as a foundational profile for the user simulator, providing a broad base of interests and inclinations.
Reflected preferences, on the other hand, enrich the conversation context by allowing the user simulator to dynamically adapt to the interaction, accordingly refine its preferences, and thoughtfully respond to the CRS.
Figure~\ref{fig:overview} illustrates the overall interaction flow of our framework.

\medskip
\noindent\textbf{General Preferences}. \label{sec:general-preferences}
To establish general preferences, we leverage a real-world user database with extensive interaction histories and informative reviews. 
These reviews provide insights into personal preferences that extend beyond simple item attributes, capturing nuanced opinions on aspects such as storyline, pacing, and emotions.
However, given that user-generated reviews often contain noise and ambiguous expressions, following \cite{kim2024pearl}, we employ ChatGPT to extract and transform each collected reviews into clear, structured binary preferences categorized into \textit{Likes} and \textit{Dislikes}.
We then partition each user's interaction history into two distinct subsets: \textit{seen} items and \textit{target} items.
The \textit{seen} items refer to those the user has previously interacted with.
In contrast, the \textit{target} set, reserved for CRS evaluation, consists exclusively of highly rated items, ensuring a reasonable basis for their use as the evaluation set.
When generating general preference, we provide ChatGPT with metadata and corresponding binary preferences derived solely from the \textit{seen} items. 
The model is then instructed to generate descriptive narratives highlighting the most representative features.
These narratives are subsequently used to initialize our simulators, each tailored to mimic a distinct instance from the user database.
Through this approach, we ensure that user simulators remain uninformed of target items while being robustly grounded in detailed general preferences. 
This grounding allows their preferences to be sufficiently generalizable to discover target items, thereby closely emulating real users.

\medskip
\noindent\textbf{Reflected Preferences}. Beyond simply articulating general preferences, real users evaluate items through the lens of their past interactions.
They tend to uncover their implicit preferences while interacting with recommendation systems, showing a dynamic and adaptable nature. 
Reflected preference functions to capture this nuanced user behavior, enabling user simulators to reflect their preferences with regard to current recommendations responsively. 
To achieve this, we categorize the items recommended by the CRS at each turn into two sets: a \textit{seen} set and an \textit{unseen} set.
For \textit{seen} items, we allow the user simulators to revisit their corresponding reviews and recalling what they liked or disliked.
For \textit{unseen} items, we prompt the user simulators to shape opinions based on their general preferences, identifying what they are expected to like or dislike.
These reflected preferences are then provided as additional input for the user’s subsequent response.
This approach enables user simulators to proactively provide feedback on both previously interacted items and newly encountered ones, consequently enriching the dialogue.

\subsection{Evaluation on Preference Elicitation}
\label{subsec:Evaluation}
Since the preference elicitation ability can be defined as \textit{"how proactively a CRS leads the conversation in a natural and engaging manner, guiding the user through discovering a diverse range of preferences to achieve a satisfactory experience"}, we consider the following key aspects:

(1) \textbf{\textit{Preference Coverage}}: evaluates how effectively CRSs discover the diverse preferences of users through the dialogue.
(2) \textbf{\textit{Proactiveness}} ~\cite{deng2024towards}:  characterizes a CRS that actively guides the conversation by making suggestions or asking relevant questions to actively uncover and clarify the user's preferences.
(3) \textbf{\textit{Coherence}}~\citet{dziri2019evaluating}: 
reflects the CRS's proficiency in maintaining fluid and natural interactions, providing contextually appropriate responses.
(4) \textbf{\textit{Personalization}}~\cite{lin2023enhancing}: 
refers to how well the system provides recommendations and information that align with the user's preferences, ensuring a satisfying interaction experience.

Based on these key aspects, we analyze CRSs both quantitatively and qualitatively.
For quantitative analysis, we measure \pc to assess how the CRS identifies each user’s target items with high coverage throughout the conversation. 
For qualitative analysis, we evaluate Proactiveness, Coherence, and Personalization to assess how effectively the CRS integrates the preference elicitation process into the conversation.

\paragraph{Quantitative Metric.}
To quantitatively measure how well the system understands user's evolving preferences and makes accurate recommendations as the conversation progresses, we propose novel metrics, \pc (PC) and \pcir (PCIR).
Specifically, PC is defined as follows:

\begin{equation}
\small
    \label{eq:PC}
    \text{PC}_t = \frac{1}{|U|} \sum_{u \in U} \cfrac{|(\bigcup_{x=1}^t P_x^u) \cap Y(u)|}{|Y(u)|}
\end{equation}
Here, $U$ denotes the set of users, $Y(u)$ is the set of target items for user $u \in \mathcal{U}$, and $P_x^u$ represents the list of items recommended to user $u$ at turn $x$. 
This metric cumulatively measures the capability of a CRS to address diverse user preferences and provide accurate recommendations.
Building on this concept, we additionally define \pcir at round t as follows:
\begin{equation}
\small
    \label{eq:PCIR}
    \textsc{PCIR}_t = \textsc{PC}_{t} - \textsc{PC}_{t-1}
\end{equation}
\textsc{PCIR$_{t}$} indicates the change of \pc between round $t-1$ and $t$.
The incremental rate of \textsc{PC} reflects how effectively the system discovers new preferences and delivers corresponding recommendations at each turn.

\paragraph{Qualitative Metric.}
To qualitatively analyze the preference elicitation ability of CRSs, following \cite{liu2023g}, we adopt an automated approach, employing an LLM (i.e., GPT-4o) as the evaluator. 
Specifically, we task the LLM with fine-grained 1-to-5 scoring rubrics with specified criteria for each rating to evaluate \proact, \coh, and \personal based on generated dialogues and each simulator’s general preferences.

\section{Experiments}
\label{sec:exp}

We conduct comprehensive experiments to demonstrate the reliability of \proposed.
Implementation of user simulation and qualitative evaluation is detailed in Appendix \ref{appendix:user_sim_appendix} and \ref{appendix:qualitative_implementation}, respectively.

\subsection{Reliability of PEPPER} \label{sec:gp_eval}
\medskip
\noindent\textbf{Target-free user simulator of PEPPER closely reflects human preferences.}
We investigate the extent to which our target-free user simulator can truly represent human preferences.
To achieve this, we structure our experiments using rating information, as it provides a clear and quantifiable indication of user preferences for items.
For comparison, we provide baseline user simulators initialized with raw reviews and binary preferences (e.g., Likes and Dislikes) to study the effectiveness of \textit{general preference} described in Section ~\ref{sec:general-preferences}.
As shown in Table~\ref{tbl:simulator}, we observe that our simulator impressively identifies high-rated items that align with its actual user ratings, achieving an accuracy of 69.5\%.
In contrast, our findings reveal that raw reviews and binary preferences are less effective in representing real user preferences.
This highlights the importance of reducing noise and ambiguity in raw reviews and modeling user preference with detailed narratives rather than simplistic binary expressions.

\begin{table}[h]
\centering
\resizebox{\linewidth}{!}{
    \begin{tabular}{lc}
    \toprule
    \textbf{User Preference Representation Types} & \textbf{Accuracy (\%)} \\
    \midrule
    Raw review &  50.6 \\
    Binary preference & 60.8 \\
    \proposed (General Preference) & 69.5 \\
    \bottomrule
    \end{tabular}
}
\caption{Evaluation results of target-free user simulator's capability to reflect human preference.}
\vspace{-0.35cm}
\label{tbl:simulator}
\end{table}

\begin{figure}[!h]
    \centering
    \includegraphics[width=1\columnwidth]{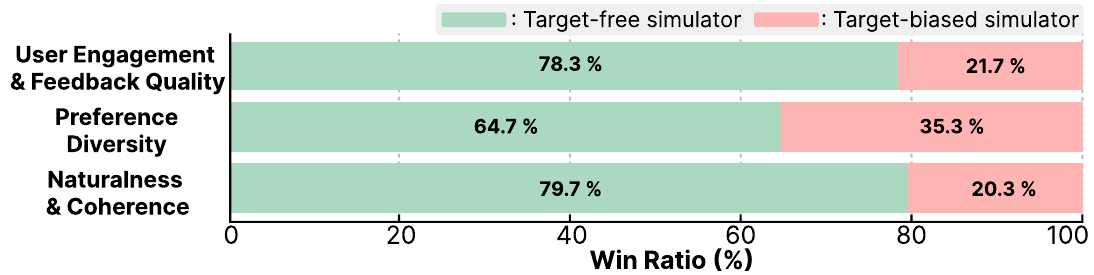}
    \caption{Human evaluation on the quality of generated dialogues from Target-free vs Target-biased simulator.
    }
    \vspace{-0.3cm}
    \label{fig:human_eval}
    \vspace{-0.2cm}
\end{figure}

\paragraph{Target-free user simulator of PEPPER closely emulates human behavior.}
To further demonstrate the efficacy of our approach, we conduct a human evaluation via Amazon Mechanical Turk (AMT). 
Specifically, we compare the quality of generated dialogues from target-biased and target-free user simulations, focusing on how effectively the user simulators provide meaningful feedback and how naturally the dialogue flows without resembling a {\textit{guessing game}}.
We compare 100 randomly sampled dialogues from both user simulations.
The results, shown in Figure ~\ref{fig:human_eval},
demonstrate that our approach achieves superior performance in capturing diverse user behaviors and maintaining a fluid dialogue progression, highlighting its effectiveness in producing realistic interactions.

\medskip
\noindent\textbf{Target-free user simulator of PEPPER mitigates bias.}
We provide a comparative analysis to further reveal the extent of bias introduced by target-biased user simulations.
Our findings in Section~\ref{sec:analysis} shows that target-biased simulations result in significant performance disparities; this limitation becomes even more evident when measured with PC.
As shown in Table~\ref{tbl:biasresults}, biased simulators significantly inflate the performance for the \textit{selected} set.
In contrast, target-free simulators demonstrate consistent PC, indicating balanced exploration across all target items. 
This suggests that target-free approach ensures unbiased simulation, providing a reliable framework for evaluating preference elicitation.
\setlength{\tabcolsep}{2pt}   
\renewcommand{\arraystretch}{1.1}
\newcommand{\GradDelta}[1]{#1}  
\newcolumntype{C}{@{\hspace{3pt}}c@{\hspace{3pt}}}
\begin{table}[h]
\centering
\footnotesize                       
\resizebox{\columnwidth}{!}{        
\begin{tabular}{p{1.2cm} C *{6}{C}} 
\toprule
\multirow{2}{*}{\textbf{Dataset}} & \multirow{2}{*}{\textbf{CRS}} &
\multicolumn{3}{c}{\textbf{Target-{biased}}} &
\multicolumn{3}{c}{\textbf{Target-free}} \\
\cmidrule(lr){3-5}\cmidrule(lr){6-8}
 & & PC$_\mathrm{sel}$ & PC$_\mathrm{res}$ & $\boldsymbol{\Delta}$ &
     PC$_\mathrm{sel}$ & PC$_\mathrm{res}$ & $\boldsymbol{\Delta}$ \\
\midrule
\multirow{4}{*}{\rotatebox{90}{\parbox[c][1.1cm][c]{2.2cm}{\centering IMDB \\ \textsuperscript{ReDial}}}}
  & KBRD   & 0.050 & 0.030 & \cellcolor{gray!20}\GradDelta{-0.020} & 0.067 & 0.062 & \cellcolor{gray!10}\GradDelta{-0.005} \\
  & BARCOR & 0.210 & 0.067 & \cellcolor{red!10}\GradDelta{-0.143} & 0.111 & 0.102 & \cellcolor{gray!10}\GradDelta{-0.009} \\
  & UniCRS & 0.372 & 0.077 & \cellcolor{red!20}\GradDelta{-0.295} & 0.078 & 0.080  & \cellcolor{gray!10}\GradDelta{+0.002} \\
  & ChatGPT& 0.880 & 0.067 & \cellcolor{red!40}\GradDelta{-0.813} & 0.125 & 0.132 & \cellcolor{gray!10}\GradDelta{+0.007} \\
  & ChatCRS& 0.873 & 0.047 & \cellcolor{red!40}\GradDelta{-0.826} & 0.129 & 0.127 & \cellcolor{gray!10}\GradDelta{-0.002} \\
  & MACRS& 0.850 & 0.072  & \cellcolor{red!40}\GradDelta{-0.778} & 0.118 & 0.120 & \cellcolor{gray!20}\GradDelta{-0.002} \\
\midrule
\multirow{4}{*}{\rotatebox{90}{\parbox[c][1.1cm][c]{2.2cm}{\centering IMDB \\ \textsuperscript{OpenDialKG}}}}
  & KBRD   & 0.063 & 0.060 & \cellcolor{red!5}\GradDelta{-0.003} & 0.098 & 0.100  & \cellcolor{gray!10}\GradDelta{+0.002} \\
  & BARCOR & 0.197 & 0.090 & \cellcolor{red!5}\GradDelta{-0.107} & 0.113 & 0.120 & \cellcolor{gray!10}\GradDelta{+0.007} \\
  & UniCRS & 0.295 & 0.102 & \cellcolor{red!20}\GradDelta{-0.193} & 0.133 & 0.165 & \cellcolor{red!5}\GradDelta{+0.032} \\
  & ChatGPT& 0.883 & 0.205 & \cellcolor{red!35}\GradDelta{-0.678} & 0.215 & 0.218 & \cellcolor{gray!10}\GradDelta{+0.007} \\
  & ChatCRS& 0.877 & 0.140  & \cellcolor{red!40}\GradDelta{-0.737} & 0.267 & 0.253 & \cellcolor{gray!20}\GradDelta{-0.014} \\
  & MACRS& 0.905 & 0.192 & \cellcolor{red!40}\GradDelta{-0.713} & 0.235 & 0.202 & \cellcolor{red!5}\GradDelta{-0.033} \\
\bottomrule
\end{tabular}}
\vspace{-0.3cm}  
\caption{Recommendation Accuracy of CRSs under target-biased and target-free user simulations. We report PC\textsubscript{selected}@50, PC\textsubscript{residual}@50, and their difference (\textbf{$\Delta$}) from 100 randomly sampled user instances.}
\vspace{-0.3cm}
\label{tbl:biasresults}
\end{table}

\noindent\textbf{Target-free user simulator of PEPPER achieves higher response diversity. }
Motivated by findings that simulators often generate repetitive, low-variety requests~\cite{yoon2024evaluating}, we compare the semantic diversity of user requests generated by PEPPER with those produced by the target-biased baseline.
Specifically, we compute pairwise embedding distances across all simulated dialogues for each simulator and averaged these distances. 
As shown in Table ~\ref{tab:user_response_diversity}, target-biased simulators generate more repetitive and less varied user requests compared to target-free user simulators, which produces more diverse utterances that better reflect authentic user behavior.
We further evaluated item-level diversity by extracting all unique items mentioned across the dialogues, providing additional evidence for the robustness of PEPPER.  
The detailed results are presented in Appendix~\ref{appendix:item_diversity}.
\begin{table}[t]
\centering
\footnotesize
\resizebox{\columnwidth}{!}{
\begin{tabular}{l l cc}
\toprule
\textbf{Model} & \textbf{Dataset} & \textbf{Target-free} $\downarrow$ & \textbf{Target-biased} $\downarrow$ \\
\midrule
\multirow{2}{*}{MACRS}
  & OpenDialKG & 0.5671 & 0.8928 \\
  & Redial     & 0.5849 & 0.8797 \\
\midrule
\multirow{2}{*}{ChatCRS}
  & OpenDialKG & 0.6337 & 0.9453 \\
  & Redial     & 0.6309 & 0.9435 \\
\midrule
\multirow{2}{*}{ChatGPT}
  & OpenDialKG & 0.6555 & 0.9041 \\
  & Redial     & 0.6595 & 0.9068 \\
\midrule
\multirow{2}{*}{BARCOR}
  & OpenDialKG & 0.8517 & 0.9758 \\
  & Redial     & 0.7233 & 0.9761 \\
\midrule
\multirow{2}{*}{UniCRS}
  & OpenDialKG & 0.8646 & 0.9806 \\
  & Redial     & 0.8586 & 0.9657 \\
\midrule
\multirow{2}{*}{KBRD}
  & OpenDialKG & 0.8979 & 0.9885 \\
  & Redial     & 0.8877 & 0.9697 \\
\bottomrule
\end{tabular}}
\caption{Quantitative Analysis of Response Diversity under Target-Free vs. Target-Biased Settings. This table reports the mean cosine similarity of seeker-generated responses. Lower values indicate higher diversity.}
\label{tab:user_response_diversity}
\end{table}

\begin{table}[h]
\centering\small
\setlength{\tabcolsep}{5pt}
\resizebox{\linewidth}{!}{
\begin{tabular}{lcc}
\toprule
\textbf{Evaluation Criteria}  & \textbf{Agreement} & \textbf{Cohen's Kappa (95\%CI)} \\
\midrule
Proactiveness       & 88.00 & 0.81 \\
Coherence & 92.00 & 0.87 \\
Personalization    & 96.00 & 0.93 \\
\bottomrule
\end{tabular}
}
\vspace{-0.2cm}
\caption{Both human evaluators and PEPPER rate the samples on a 1–5 Likert scale. We report the agreement rate and Cohen’s Kappa between PEPPER and human.}
\vspace{-0.2cm}
\label{tbl:qualitativehuman}
\end{table}

\begin{table*}[ht]
\centering
\setlength{\tabcolsep}{4pt}
\resizebox{0.99\linewidth}{!}
{
    \begin{tabular}{cccccccccc}
    \toprule
    \multirow{2.5}{*}{\textbf{Dataset}} & \multirow{2.5}{*}{\textbf{CRS}} & \multicolumn{8}{c}{\textbf{Evaluation Metric}} \\ \cmidrule(lr){3-10}
    & & $\bm{\mathrm{PC}_{20}@5}$ & {$\bm{\mathrm{PC}_{20}@10}$} & {$\bm{\mathrm{PC}_{20}@20}$} & {$\bm{\mathrm{PC}_{20}@50}$}
    & {$\bm{\mathrm{Recall}@5}$} & {$\bm{\mathrm{Recall}@10}$} & {$\bm{\mathrm{Recall}@20}$} & {$\bm{\mathrm{Recall}@50}$} \\
    \midrule 				
    \multirow{4}{*}{\rotatebox{90}{\parbox{3cm}{\centering IMDB \\ \textsuperscript{ReDial}}}}
    & \kbrd 
    & 0.0081 & 0.0127 & 0.0194 & 0.0477
    & \uline{0.0066} & \uline{0.0120} & \uline{0.0178} & 0.0353 \\
    
    & \barcor
    & 0.0155 & 0.0307 & 0.0472 & 0.0915
    & \textbf{0.0072} & \textbf{0.0128} & \textbf{0.0225} & \textbf{0.0525} \\
    
    & \unicrs 
    & 0.0097 & 0.0186 & 0.0447 & {0.0905}			
    & 0.0035 & 0.0052 & {0.0177} & \uline{0.0375} \\			
    
    & \chatgpt 
    & \uline{0.0334} & \uline{0.0495} & \uline{0.0671} & \uline{0.1041}				
    & {0.0011} & {0.0035} & 0.0053 & 0.0135 \\		

    & \chatcrs
    & \textbf{0.0339} & \textbf{0.0547} & \textbf{0.0792} & \textbf{0.1266}				
    & {0.0007} & {0.0024} & {0.0065} & {0.0169} \\
    
    & \macrs
    & {0.0193} & {0.0351} & {0.0586} & {0.1031}				
    & {0.0021} & {0.0025} & {0.0032} & {0.0160} \\
    \midrule   
    \multirow{4}{*}{\rotatebox{90}{\parbox{3cm}{\centering IMDB \\ \textsuperscript{OpenDialKG}}}}
    
    
    
    
    

    &\kbrd 
    & 0.0114 & 0.0256 & 0.0465 & 0.1042			
    & {0.0037} & 0.0069 & 0.0141 & 0.0410 \\	
    
    &\barcor 
    & 0.0074 & 0.0177 & 0.0488 & 0.1119	
    & 0.0025 & {0.0064} & {0.0196} & \textbf{0.0561} \\							
    &\unicrs
    & {0.0245} & {0.0397} & {0.0681} & {0.1542}	
    & {0.0044} & {0.0075} & {0.0121} & {0.0252} \\
    
    &\chatgpt 
    & \textbf{0.0685} & \uline{0.0937} & \uline{0.1410} & \uline{0.2290}			
    & \textbf{0.0083} & \textbf{0.0150} & \uline{0.0203} & 0.0423 \\
    
    & ChatCRS 
    & \uline{0.0665} & \textbf{0.0943} & \textbf{0.1437} & \textbf{0.2385}	
    & {0.0042} & {0.0093} & {0.0189} & \uline{0.0466} \\
    
    & MACRS 
    & 0.0521 & {0.0856} & {0.1243} & {0.2127}				
    & \uline{0.0056} & \uline{0.0125} & \textbf{0.0211} & {0.0364} \\
    \bottomrule
    \end{tabular}
}
\vspace{-0.2cm}
\caption{Evaluation of CRSs under our evaluation protocol.
We report \pc and Avg.Recall across 20 conversation turns to evaluate both the preference elicitation and recommendation accuracy of CRSs.}
\vspace{-0.3cm}
\label{tbl:mainresults}
\end{table*}

\medskip
\noindent\textbf{{Qualitative measure of PEPPER aligns with human judgement.}}
To further validate the reliability of the qualitative metric in PEPPER, we conduct a meta-evaluation to verify its alignment with human judgments.
Specifically, we collect human ratings for a total of 100 samples.
Each response is evaluated by human annotators based on the same rubric for \textit{Proactiveness}, \textit{Coherence}, and \textit{Personalization}.
We then compute the percentage of agreement and Randolph's Kappa between the human ratings and the scores produced by PEPPER. 
From the results in Table~\ref{tbl:qualitativehuman}, the agreement rates between PEPPER and human annotators reach 88\% for Proactiveness, 92\% for Coherence, and 96\% for Personalization, with corresponding Cohen’s Kappa of 0.81, 0.87, and 0.93, respectively, indicating a strong alignment between PEPPER and human assessments.

\subsection{CRS Evaluation with PEPPER}
Leveraging PEPPER, we evaluate and analyze the performance of existing CRS baselines with both quantitative and qualitative measures.  

\medskip
\noindent\textbf{Quantitative Evaluation.}
As shown in Table~\ref{tbl:mainresults}, LLM-based models demonstrate superior performance compared to supervised models when evaluated using PC. 
This advantage can be attributed to their advanced conversational capabilities, which enable more effective preference elicitation through natural language interactions with users. 
However, when evaluated using Recall as the performance metric, this superiority is no longer evident. 
Notably, \kbrd and \barcor exhibit higher Recall performance than their LLM-based counterparts in {IMDB\textsubscript{Redial}} dataset. 
This further strengthens the findings in Section \ref{sec:recall_fails}, indicating that while Recall is effective for measuring per-turn target item accuracy, it fails to assess preference elicitation at the dialogue level, which is better reflected by PC.
We also provide a reproducibility study using open-source models, where PEPPER consistently yields comparable results, as detailed in Appendix~\ref{appendix:opensource}.

\begin{figure}[t]
    \centering
    \includegraphics[width=1\columnwidth]{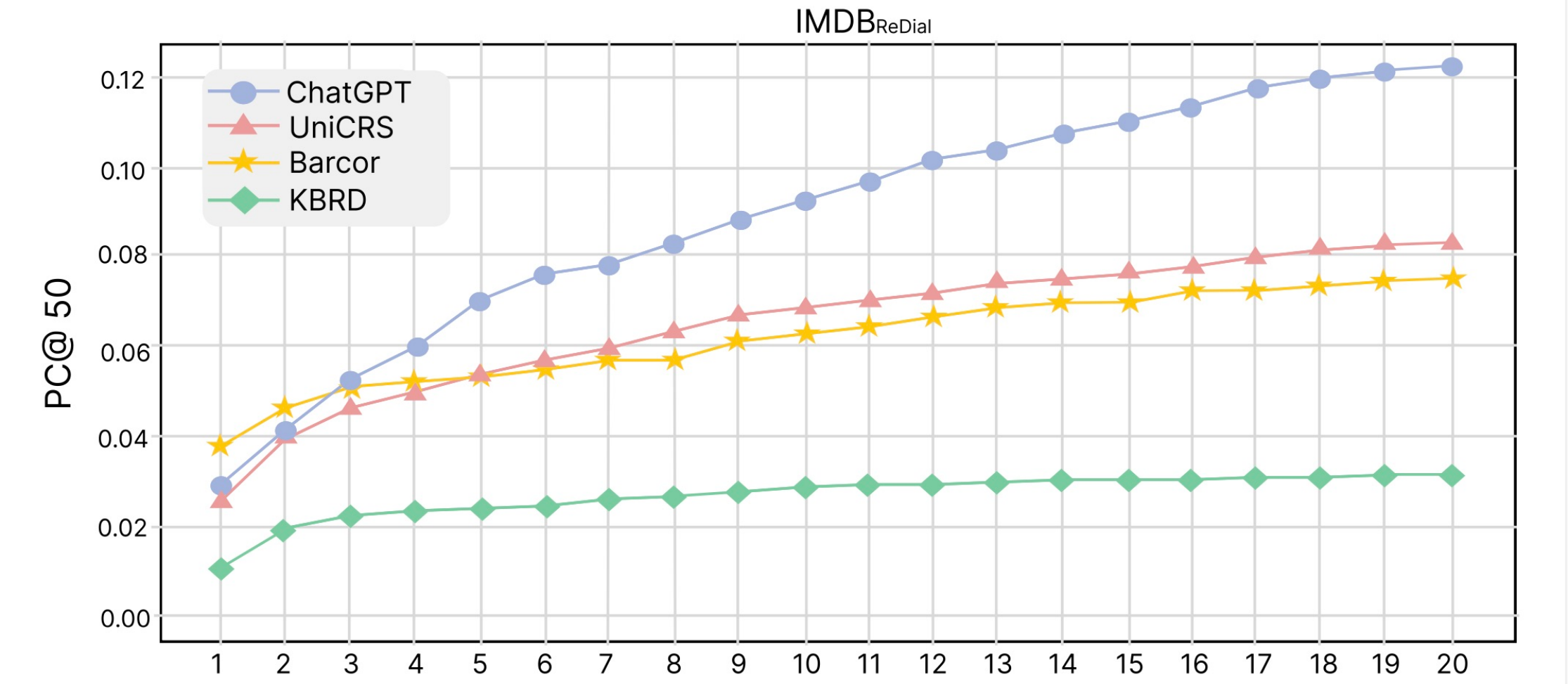}
    \caption{PC values of the CRSs for every turn $t$ in the IMDB\textsubscript{ReDial} dataset.}
    \label{fig:PC@50} 
    \vspace{-0.3cm}
\end{figure}

To gain deeper insights into how preference elicitation unfolds over time, we analyze PC at each turn of the dialogue.
As shown in Figure~\ref{fig:PC@50}, \chatgpt maintains a consistently upward trend in PC over turns, suggesting a sustained effort to explore user preferences incrementally rather than relying solely on revealed information.
In contrast, supervised baselines exhibit slower PC growth, reflecting more reactive interactions.
These trends are further supported by the PCIR scores in Table~\ref{tbl:elicitation}, where LLM-based CRSs generally achieve higher performance,  highlighting their proactive exploration of evolving user preferences and the ability to adapt recommendations throughout the dialogue.

\begin{table}[t]
\centering
\setlength{\tabcolsep}{1mm}
\resizebox{1\linewidth}{!}
    {
    \begin{tabular}{cccccc}
    \toprule
    & \multirow{2.5}{*}{\textbf{CRS}} & \multicolumn{4}{c}{\textbf{Evaluation Metric}} \\
    \cmidrule(lr){3-6}
     & & \textbf{PCIR}$\bm{_{avg}}$ & \textbf{\proact} & \textbf{\coh} & \textbf{\personal} \\
     \midrule
     \multirow{4}{*}{\rotatebox{90}{\parbox{3cm}{\centering IMDB \\ \textsuperscript{ReDial}}}}  &

     \kbrd &{0.0019}&{1.10}&{1.06}&{1.2}\\
     & \barcor & {0.0019}&{1.70}&{1.83}&{1.62}\\
     & \unicrs & {0.0030}& {1.26}&{1.41}&{1.25}\\
     & \chatgpt & 0.0043&\uline{3.79}&\uline{4.55}&\textbf{4.00}\\
     & \chatcrs & \textbf{0.0059} & \textbf{4.18} & \textbf{4.93} & \uline{3.98} \\
     & \macrs & \uline{0.0045} & {3.68} & {4.08} & {3.36} \\
     
     
     \midrule
     \multirow{4}{*}{\rotatebox{90}{\parbox{3cm}{\centering IMDB \\ \textsuperscript{OpenDialKG}}}} &
     
     \kbrd &{0.0016} &{1.74}&{1.00}&{1.21}\\
     & \barcor &{0.0030} &{1.51}&{1.61}&{1.30}\\
     & \unicrs & {0.0050}&{1.11}&{1.08}&{1.2}\\
     & \chatgpt & 0.0081&\uline{3.95}&\uline{4.87}&\textbf{3.9}\\
     & ChatCRS & \textbf{0.0102} & \textbf{4.16} & \textbf{4.90} & \uline{3.83} \\
     & MACRS & \uline{0.0090} & {3.77} & {4.20} & {3.46} \\
     
    \bottomrule
    \end{tabular}
}
\vspace{-0.3cm}
\caption{Comparison on preference elicitation performances of the CRSs.
The $\text{PCIR}_{avg}$ denotes the average PCIR value per turn across the entire conversation.}
\vspace{-0.6cm}
\label{tbl:elicitation}
\end{table}

\medskip
\noindent\textbf{Qualitative Evaluation.}
Table~\ref{tbl:elicitation} shows that LLM-based CRSs significantly outperform supervised models in terms of \textit{Proactiveness}, \textit{Coherence}, and \textit{Personalization}. 
These results, also supported by Figure~\ref{fig:PC@50}, show that LLM-based approaches achieve higher PC scores, demonstrating their ability to effectively capture context shifts throughout the dialogue and seamlessly adapt to user feedback.
Comparing different LLM-based CRSs, we observe that
\chatcrs attains the highest level of \textit{Proactiveness}, which can be attributed to its goal-guidance module that actively drives the conversation toward preference elicitation rather than passively waiting for user feedback. 
By contrast, \macrs uses a multi-agent framework to diversify conversational strategies yet restricts the action space mainly to asking, recommending, and chit-chatting.
This design choice, while promoting structured interactions, may limit its ability to engage in more flexible or nuanced proactive behaviors.

\section{Conclusion}
\label{sec:conclusion}
In this work, we propose \proposed, a novel evaluation protocol that comprehensively assesses both preference elicitation and recommendation accuracy in CRSs.
PEPPER incorporates target-free user simulators, along with both quantitative and qualitative metrics, targeting four distinct aspects of the preference elicitation process.
Through extensive experiments, we demonstrate the effectiveness of \proposed, offering valuable insights into the limitations of existing CRS evaluation protocols. 
\section*{Limitations}
\label{sec:limitation}
While our study offers valuable insights into evaluating preference elicitation in CRS, it is not without limitations.
One limitation is that our experiments are conducted in the movie and e-commerce domain, where user preferences are well-articulated through reviews.
This setting allows us to simulate nuanced behaviors in a controlled environment, but generalizing to other domains remains an open challenge.
We believe the design of our simulator is domain-agnostic and can be adapted to new settings, though further validation is required.

Another limitation lies in our reliance on proprietary LLM (GPT-4o-mini) for both simulation and evaluation, which may introduce generation patterns not fully representative of other models~\citep{seo-etal-2025-mt}.
To reduce this concern, we provide additional results using LLaMA-3.1-8B-Instruct and Mistral-7b-Instruct, confirming the robustness of our framework across different architectures.

A further limitation is that while PEPPER presents new evaluation metrics and perspectives for understanding CRS behaviors, it does not explore methods for improving CRS models themselves.
The focus of this work is to analyze how existing systems perform in eliciting user preferences through dialogue.
Future work could build on these insights to develop CRS architectures that better support preference elicitation and adapt more effectively to evolving user needs.

\section*{Ethical Consideration}
\label{sec:ethical}
Text generated by LLMs may contain content that is harmful or biased, and have the potential risk of hallucination~\citep{kim-etal-2024-verifiner,Seo2024UnveilingIT}. 
However, in our research, we take several steps to minimize these risks. The source dataset, IMDb Movies, is publicly available under the CC0 Public Domain license and includes human-annotated data. 
Additionally, we manually inspect and filter the dialogues generated through user–CRS interactions to eliminate toxic, offensive, or biased language.
For human evaluation, we recruit three independent annotators per unit task via Amazon Mechanical Turk (AMT), ensuring fair compensation. 
Each annotator is paid \$0.15 per task. 
The textual content presented in this paper contains no personally identifiable information and poses no risk of re-identifying individuals or groups.

\section*{Acknowledgement}
This work was supported by the IITP grants funded by the Korea government (MSIT) (No. RS-2020-II201361; RS-2024-00457882, AI Research Hub Project), and the KBSI (National research Facilities and Equipment Center) grant funded by the Korea government (MSIT) (No. RS-2024-00403860).

\bibliography{custom}

\appendix
\label{sec:appendix}
\clearpage
\section{Appendix}
\subsection{{Dataset}}\label{appendix:dataset}

IMDB is a comprehensive movie database that features extensive user profiles with rich interaction histories and detailed reviews.
\redial is a CRS dataset focused on movie recommendations, created using crowd-sourced dialogues through Amazon Mechanical Turk (AMT). 
\opendialkg is also a CRS dataset with a broader range of domains, including movies, sports, books and music.
However, in this study, we focus on the movie domain due to its accessibility and prominence in CRS research ~\cite{jannach2021survey}.
We have manually enriched the \opendialkg dataset by collecting movie plots from the \imdb website, as it does not provide movie plots in its metadata.
To ensure reliable preference modeling, we also excluded users with fewer than 10 interactions.
The statistics of the processed \imdb user dataset are summarized in Table~\ref{tbl:dataset}.
\begin{table}[ht]
\centering
\setlength{\tabcolsep}{1mm} 
\resizebox{0.95\linewidth}{!}{
    \begin{tabular}{lcc}
    \toprule
    \multicolumn{1}{c}{\textbf{Dataset}} & \textbf{\#Users} & \textbf{\#Interaction Histories} \\
    \midrule
    IMDB\textsubscript{ReDial} & 3,306 & 66,075 \\
    IMDB\textsubscript{OpenDialKG} & 2,666 & 47,337 \\
    \bottomrule
    \end{tabular}
}
\caption{Statistics of processed datasets.}
\label{tbl:dataset}
\end{table}


\subsection{CRS Baselines}\label{appendix:crs}
We conduct a comparative analysis of six representative CRSs, encompassing both supervised approaches—\kbrd~\cite{chen2019towards}, \barcor~\cite{wang2022barcor}, \unicrs~\cite{wang2022towards}—and LLM-based methods: \chatgpt, \chatcrs~\cite{li2025chatcrs}, \macrs~\cite{fang2024multi}.
For supervised CRS model implementation, we adhere to \cite{wang2023rethinking}.
For LLM-based approaches, we employ gpt-4o-mini~\cite{ouyang2022training} as the conversation module.
We also integrate a recommender module using the \texttt{text-embedding-ada-002} model~\cite{neelakantan2022text} for LLM-based CRS to constrain the output space of LLM based methods, as they tend to generate items that are beyond the scope of evaluation datasets.
Inspired by \cite{friedman2023leveraging, zhang2024generative}, we introduce an item interface, enabling user simulators to interact with the current recommendations. 
This approach more closely mirrors real-world scenarios, where users actively engage with recommendations and provide implicit feedback, facilitating the dynamic refinement of their preferences. 
Recommendations are retrieved using each CRS’s specific retrieval model.
Once retrieved, the items are manually augmented with corresponding plots and incorporated into the reflection generation prompts of our user simulators.
\begin{itemize}[noitemsep]
    \item \textbf{\kbrd}: enhances the semantic understanding of entities mentioned in conversation history by bridging the recommendation module and transformer-based conversation module through knowledge propagation.
    \item \textbf{\barcor}: presents a unified framework based on BART \cite{lewis2019bart} that integrates both recommendation and response generation tasks into a single model.
    \item \textbf{\unicrs}: proposes a unified framework based on DialoGPT ~\cite{zhang2019dialogpt} that incorporates a semantic fusion module and knowledge-enhanced prompt learning to improve the association between dialogue history and knowledge graphs. 
    \item \textbf{\chatgpt}: is an LLM that exploits in-context learning to adapt flexibly across tasks, making it widely effective for conversational and recommendation scenarios.
    \item \textbf{\chatcrs}: introduces a multi-agent CRS framework with a knowledge-retrieval agent and a goal-planning agent that leverage external knowledge and goal guidance.
    \item \textbf{\macrs}: presents a CRS that combines multi-agent action planning with user feedback–aware reflection to adapt strategies and enhance recommendations.
\end{itemize}


\subsection{Target-biased User Simulation} \label{appendix:biasuser}
We use \texttt{gpt-4o-mini} as the backbone language model to simulate the target-biased user simulator.
Following prior work~\cite{zhu2024reliable, huang2024concept, zhu2025llm, wang2023rethinking}, the user simulator is modeled with target item attributes, including genres, directors, stars, and plot summaries, with the item title intentionally excluded.
Each dialogue is simulated for up to 20 turns, allowing sufficient interaction for preference elicitation. We evaluate the performance using 100 sampled user instances from each dataset.

\subsection{Target-free User Simulation} \label{appendix:user_sim_appendix}
\subsubsection{Interaction Environment}
Our interaction environment comprises two generative agents: a target-free user simulator and a CRS. These agents engage through a dialogue interface and an item interface.
The dialogue interface bridges communication between the user and the CRS, while the item interface presents top-$K$ recommendations predicted by the CRS at each turn, along with their metadata (\textit{i.e.,} movie plots).
By incorporating the item interface, we closely emulate real-world scenarios where users can access detailed information about the recommended items.

For user simulation, we start by extracting the most representative preferences from a user's raw reviews and categorize them into \textit{Likes} and \textit{Dislikes}.
These preferences are then transformed into descriptive narratives, depicting the general preferences of the user simulator.
Next, the user simulator initiates a entirely new conversation by requesting recommendations that align with its general preferences. 
In response, the CRS generates an utterance and presents the top-$K$ item suggestions through the item interface.
As interactions continue, the user simulator not only communicates with the recommender but also engages with the item interface by carefully examining each suggested item. 
For previously interacted items (i.e., seen), it retrieves past reviews, while for newly encountered items (i.e., unseen), it shapes opinions based on its general preferences.
This dual engagement allows the simulator to elicit its own preferences and provide detailed feedback during subsequent interactions, thereby enriching the dialogue to better align with the user's interests and facilitating the discovery of relevant items.

\label{appendix:freeuser}
\subsubsection{Implementation Detail}
We conduct experiments using 500 user simulators for each dataset.
We adopt \texttt{gpt-4o-mini} for our target-free user simulations, comprising (1) preference extraction, (2) general preference generation, (3) reflected preference generation, and (4) response generation.
We leverage zero-shot prompting to guide the model through each process.
To maintain consistent and deterministic outputs, we fix the temperature parameter for user simulation at 0. 
The number of items presented in the item interface is set to 4, and each simulated dialogue continues up to 20 turns.

\subsubsection{Evaluating Simulator's Capability to Represent Human Preference}\label{appendix:rating}
In Section \ref{sec:gp_eval}, we provide a study to evaluate how closely the proposed target-free user simulator reflects real human preferences.
The experiment is conducted as follows: first, a user simulator takes a pair of target items rated by its corresponding user.
Then, we instruct the simulator to select the item that aligns more closely with its general preferences.
Afterward, we assess the simulator's ability to correctly identify the item with the higher rating based on the actual user scores.

\begin{table}[h]
\centering
\footnotesize
\resizebox{\columnwidth}{!}{
\begin{tabular}{l l cc}
\toprule
\textbf{Model} & \textbf{Dataset} & \textbf{Target-free} & \textbf{Target-biased} \\
\midrule
\multirow{2}{*}{ChatGPT}
  & OpenDialKG & 2326 & 969 \\
  & Redial     & 2138 & 897 \\
\midrule
\multirow{2}{*}{UniCRS}
  & OpenDialKG & 141 & 55 \\
  & Redial     & 117 & 59 \\
\midrule
\multirow{2}{*}{BARCOR}
  & OpenDialKG & 262 & 151 \\
  & Redial     & 1180 & 37 \\
\midrule
\multirow{2}{*}{ChatCRS}
  & OpenDialKG & 1976 & 544 \\
  & Redial     & 1821 & 444 \\
\midrule
\multirow{2}{*}{KBRD}
  & OpenDialKG & 91 & 25 \\
  & Redial     & 54 & 27 \\
\midrule
\multirow{2}{*}{MACRS}
  & OpenDialKG & 1431 & 691 \\
  & Redial     & 1310 & 553 \\
\bottomrule
\end{tabular}}
\caption{Quantitative Analysis of Response Diversity under Target-Free vs. Target-Biased Settings. This table reports the item-level diversity measured by the number of unique items mentioned in simulated dialogues.}
\label{tab:item_diversity}
\end{table}
\begin{table}[ht]
\centering
\resizebox{0.95\linewidth}{!}
    {
    \begin{tabular}{cccccc}
    \toprule
    & \textbf{CRS} & \textbf{\small PC\textsubscript{15}@5} & \textbf{\small PC\textsubscript{15}@10} & \textbf{\small PC\textsubscript{15}@20} & \textbf{\small PC\textsubscript{15}@50} \\
    \midrule 
     \cmidrule(lr){3-6}
     \multirow{4}{*}{\rotatebox{90}{\makecell{{PEPPER} \\
     {\textsuperscript{{Llama}}}}}} & 
     
     \kbrd & {0.0050} & {0.0091} & {0.0320} & {0.0670}\\
     
     & \barcor & {0.0167} & {0.0207} & {0.0498} & {0.0993}\\
     
     & \unicrs & \uline{0.0233} & \uline{0.0350} & \uline{0.0617} & \uline{0.1022}\\
     
     & \chatgpt & \textbf{0.0287} & \textbf{0.0545} & \textbf{0.0877} & \textbf{0.1829}\\
     \midrule
     \cmidrule(lr){3-6}
     
    \multirow{4}{*}{\rotatebox{90}{\makecell{{PEPPER} \\
     {\textsuperscript{{Mistral}}}}}} & 
     
     \kbrd & {0.015} & {0.041} & {0.047} & {0.106}\\
     
     & \barcor & {0.011} & {0.024} & {0.059} & {0.136}\\
     
     & \unicrs & \uline{0.035} & \uline{0.052} & \uline{0.091} & \uline{0.173}\\
     
     & \chatgpt & \textbf{{0.079}} & \textbf{{0.118} }& \textbf{{0.156}} & \textbf{0.236}\\
     
    \bottomrule
    \end{tabular}
}
\caption{Recommendation Accuracy of CRSs using Target-free User Simulation with
Open-source LLMs.}
\label{tbl:openusersim}
\end{table}

\subsubsection{Item Diversity Analysis}\label{appendix:item_diversity}
We analyzed item-level diversity by extracting all item entities mentioned throughout the dialogues and counting the unique items referenced by each simulator. 
The results in Table ~\ref{tab:item_diversity} demonstrate that PEPPER covers a substantially wider range of items across dialogues, whereas target-biased simulators concentrate on a narrow subset of entities, resulting in significantly reduced item diversity.

\subsubsection{Target-free User Simulation with Open-source LLMs}\label{appendix:opensource}
We verify the reproducibility of \proposed through experiments using Llama-3.1-8B-Instruct and Mistral-7B-Instruct as the backbones for our target-free user simulators.
The experiments involve 100 user samples from IMDB\textsubscript{OpenDialKG}, with each conversation simulated for up to 15 turns. 
The results, presented in Table ~\ref{tbl:openusersim}, reveal that PEPPER shows consistent evaluation performance across different CRSs. 
These findings validate not only the reproducibility of our framework with open-source models but also its effectiveness for CRS evaluation.


\subsubsection{Generalizability of PEPPER}
To demonstrate PEPPER's domain robustness, we conducted additional experiments in the e-commerce domain, specifically using the Amazon Electronics dataset~\cite{hou2024bridging}.
We followed the same simulation setup and seen/unseen split protocol as in the movie domain to ensure a fair comparison across domains.
Based on the results in Table ~\ref{tbl:biascomputer}, we report two key findings: \textbf{ (1) PEPPER maintains robust simulation quality and evaluation reliability across domains.}
As shown in the Residual-vs-Selection evaluation results, PEPPER continues to outperform target-biased simulators in providing unbiased and informative simulations, even in the e-commerce setting. 
This suggests that PEPPER is generalizable and effective across domains, making it a reliable tool for evaluating CRS in both familiar and novel contexts.
\textbf{(2) Current CRS models exhibit significant domain-dependent performance degradation.}
We attribute this to the lack of domain-specific knowledge in current CRS. 
While many CRS systems implicitly benefit from LLMs’ rich parametric knowledge in popular domains like movies, their performance degrades in less familiar domains such as e-commerce. This is consistent with prior work ~\cite{liu2023conversational,cao2024aligning, peng2024ecellm} that highlight the challenges of domain transfer and the limitations of LLMs’ internal knowledge in handling diverse item spaces. 
We believe this result guides a crucial future direction: how to equip CRS with external or domain-adaptive knowledge sources so they can better understand and recommend items across various domains.
\setlength{\tabcolsep}{2pt}   
\renewcommand{\arraystretch}{1.1}
\newcolumntype{C}{@{\hspace{3pt}}c@{\hspace{3pt}}}
\begin{table}[h]
\centering
\footnotesize                       
\resizebox{\columnwidth}{!}{        
\begin{tabular}{p{1.2cm} C *{6}{C}} 
\toprule
\multirow{2}{*}{\textbf{Dataset}} & \multirow{2}{*}{\textbf{CRS}} &
\multicolumn{3}{c}{\textbf{Target-{biased}}} &
\multicolumn{3}{c}{\textbf{Target-free}} \\
\cmidrule(lr){3-5}\cmidrule(lr){6-8}
 & & PC$_\mathrm{sel}$ & PC$_\mathrm{res}$ & $\boldsymbol{\Delta}$ &
     PC$_\mathrm{sel}$ & PC$_\mathrm{res}$ & $\boldsymbol{\Delta}$ \\
\midrule
\multirow{4}{*}{\makecell{Amazon\\\textsubscript{Electronics}}}
  & ChatGPT& 0.147 & 0.009 & \cellcolor{red!25}\GradDelta{-0.138} & 0.095 & 0.071 & \cellcolor{red!5}\GradDelta{-0.024} \\
  & ChatCRS& 0.143 & 0.009 & \cellcolor{red!25}\GradDelta{-0.134} & 0.104 & 0.075 & \cellcolor{red!5}\GradDelta{-0.029} \\
  & MACRS& 0.147 & 0.025  & \cellcolor{red!25}\GradDelta{-0.122} & 0.106 & 0.105 & \cellcolor{gray!5}\GradDelta{-0.001} \\
\bottomrule
\end{tabular}}
\vspace{-0.3cm}  
\caption{Recommendation Accuracy of CRSs under target-biased and target-free user simulations. We report PC\textsubscript{selected}@50, PC\textsubscript{residual}@50, and their difference (\textbf{$\Delta$}) from 100 randomly sampled user instances.}
\vspace{-0.3cm}
\label{tbl:biascomputer}
\end{table}

\subsubsection{Results with Additional Metrics}
To provide a more comprehensive evaluation beyond Recall, we report additional metrics—Precision, NDCG, and MRR—as shown in Table~\ref{tbl:metrictable}.
From the results, we observe that the relative ranking of the CRS baselines remains consistent across all metrics. 
This result shows that PEPPER supports robust and fair evaluation across diverse recommendation accuracy metrics.

\begin{table*}[ht]
\centering
\setlength{\tabcolsep}{4pt}
\resizebox{0.99\linewidth}{!}
{
    \begin{tabular}{cccccccccc}
    \toprule
    \multirow{2.5}{*}{\textbf{Dataset}} & \multirow{2.5}{*}{\textbf{CRS}} & \multicolumn{8}{c}{\textbf{Evaluation Metric}} \\ \cmidrule(lr){3-10}
    & & $\bm{\mathrm{PC}_{20}@20}$ & {$\bm{\mathrm{PC}_{20}@50}$} & {$\bm{\mathrm{Recall}@20}$} & {$\bm{\mathrm{Recall}@50}$}
    & {$\bm{\mathrm{NDCG}@20}$} & {$\bm{\mathrm{NDCG}@50}$} & {$\bm{\mathrm{MRR}@20}$} & {$\bm{\mathrm{MRR}@50}$} \\
    \midrule 				
    \multirow{4}{*}{\rotatebox{90}{\parbox{3cm}{\centering IMDB \\ \textsuperscript{ReDial}}}} &
    \kbrd 
    & 0.019 & 0.048 & \uline{0.018} & 0.031
    & 0.015 & 0.024 & 0.008 & 0.009 \\ 	
    
    & \barcor
    & 0.047 & 0.092 & \textbf{0.022} & \textbf{0.053}
    & 0.028 & 0.041 & 0.010 & 0.012 \\ 	
    
    &\unicrs 
    & 0.045 & 0.091 & {0.018} & \uline{0.038}			
    & 0.021 & 0.035`` & {0.006} & 0.008 \\				
    
    &\chatgpt 
    & \uline{0.067} & \uline{0.104} & 0.005 & 0.014		
    & \uline{0.047} & \uline{0.058} & \uline{0.021} & \uline{0.022} \\	
    
    & ChatCRS
    & \textbf{0.079} & \textbf{0.127} & 0.007 & 0.017				
    & \textbf{0.054} & \textbf{0.069} & \textbf{0.025} & \textbf{0.027} \\
    
    & MACRS
    & 0.059 & 0.103 & 0.003 & 0.016				
    & {0.035} & {0.047} & 0.014 & 0.015 \\
    \midrule   
    \multirow{4}{*}{\rotatebox{90}{\parbox{3cm}{\centering IMDB \\ \textsuperscript{OpenDialKG}}}} 
    &\kbrd 
    & 0.047 & 0.104 & {0.014} & 0.041
    & 0.025 & 0.042 & 0.010 & 0.012 \\		
    
    &\barcor 
    & 0.049 & 0.112 & 0.019 & \textbf{0.056}	
    & 0.021 & 0.039 & 0.006 & 0.008 \\	
    
    &\unicrs
    & 0.068 & 0.154 & 0.012 & 0.025		
    & 0.042 & 0.067 & 0.018 & 0.020 \\			
    
    &\chatgpt 
    & \uline{0.141} & \uline{0.229} & \uline{0.020} & {0.042}			
    & \uline{0.099} & \uline{0.124} & \uline{0.049} & \uline{0.052} \\
    
    & ChatCRS
    & \textbf{0.144} & \textbf{0.239} & {0.019} & \uline{0.047}
    & \textbf{0.102} & \textbf{0.129} & \textbf{0.050} & \textbf{0.053 }\\
    
    & MACRS
    & {0.124} & {0.213} & \textbf{0.021} & 0.036				
    & {0.080} & {0.105} & {0.038} & {0.041} \\
    \bottomrule
    \end{tabular}
}
\vspace{-0.2cm}
\caption{Evaluation of CRSs under our evaluation protocol.
We report \pc, Avg.Recall, Avg.NDCG, and Avg.MRR across 20 conversation turns to evaluate both the preference elicitation and recommendation accuracy of CRSs.}
\vspace{-0.3cm}
\label{tbl:metrictable}
\end{table*}

\subsubsection{Impact of Item Quantity in Item Interface}
We explore whether changing the number of items in the item interface influences the quality of user-CRS interactions, as having more items allows the user simulator to better generate its reflected preferences.
We conduct experiments using 100 user simulators, with the number of items set to 0, 4, 7, and 10, where 0 is the setting in which preference reflection is excluded.
Each dialogue is simulated for 15 turns, and the results are shown in Table~\ref{tbl:itemnumber}. 

We observe a significant performance gap when the preference reflection process is excluded from the interaction, indicating its critical role in enhancing the quality of user-CRS interactions. 
However, when preference reflection is included, we observe that increasing the item count has no measurable impact on the interactions.
We attribute this to the behavior of our user simulators, which tend to prioritize reflecting preferences for the most relevant recommendations rather than engaging with all available options. 
In fact, some CRSs, such as UniCRS, exhibit a slight decrease in performance as the item count increases. 
This indicates that simply adding more items may instead introduce noise into the interaction process.
\begin{table}[ht]
\centering
\resizebox{0.93\linewidth}{!}
    {
    \begin{tabular}{cccccc}
    \toprule
    & \multirow{2.5}{*}{\textbf{CRS}} & \multicolumn{4}{c}{\textbf{\# of items}} \\
    \cmidrule(lr){3-6}
     & & \textbf{0 }& \textbf{4} & \textbf{7} & \textbf{10} \\
     \midrule 
     \multirow{4}{*}{\rotatebox{90}{\makecell{{IMDB} \\{\textsuperscript{{ReDial}}}}}} 
     
     & \kbrd    &{0.0199} &{0.0121}   &{0.0138}   &{0.0129} \\
     & \barcor  &{0.0715} &{0.0825}         &{0.0873}   &\uline{0.0842} \\
     & \unicrs  & \uline{0.0860} &\uline{0.0938}   &\uline{0.0936}        &{0.0772}\\
     & \chatgpt & \textbf{0.1038} &\textbf{0.1130}  &\textbf{0.1039}  &\textbf{0.1187} \\
     
     \midrule
     \multirow{4}{*}{\rotatebox{90}{\makecell{{IMDB} \\{\textsuperscript{{OpenDialKG}}}}}} 
     
     & \kbrd    &{0.1060} &{0.0845}        &{0.0737}        &{0.0662} \\
     & \barcor  &{0.0817} &{0.0968}        &{0.1043}        &{0.1005} \\
     & \unicrs  & \uline{0.1275} & \uline{0.1485}  & \uline{0.1410}  & \uline{0.1278} \\
     & \chatgpt & \textbf{0.1865} &\textbf{0.2262} &\textbf{0.2243} &\textbf{0.2033}\\
     
    \bottomrule
    \end{tabular}
}
\vspace{-0.2cm}
\caption{CRS performance in user simulations with different numbers of items presented in the item interface. We assess $PC@_{50}$ for evaluation.}
\label{tbl:itemnumber}
\vspace{-0.35cm}
\end{table}

\begin{table}[h]
\centering
\resizebox{\linewidth}{!}{%
    \begin{tabular}{p{\linewidth}}
    \toprule
    {\textbf{Ground Truth (Target Items)}: "Ocean's Eleven", "Armageddon", ...}  \\ 
    \midrule
    \rowcolor{gray!15}
    {\textbf{Dialogue Context}} \\
    \ldots \\
    {\textbf{Recommender}: I would recommend the movie ``Ocean's Eleven.'' This film features a group of charismatic and intelligent characters who plan and execute a heist in a stylish and entertaining manner.} \\
    {\textbf{User: } Ocean's Eleven seems like a promising recommendation with its cool characters and entertaining heist plot. I appreciate the potential for a stylish and engaging storyline...} \\
    \ldots \\
    \midrule
    \rowcolor{gray!15} \textbf{Item Interface} \\
    1. [\textit{Ocean's Eleven}] Danny Ocean gathers a group of his World War II compatriots to pull off the ultimate Las Vegas heist \ldots \\
    2. [\textit{Inside Out}] An agoraphobic must give up his sheltered life and venture outside after a series of personal and financial problems. \ldots \\
    3. \ldots \\
    \midrule
    \rowcolor{gray!15} \textbf{User's General Preferences} \\
    You enjoy vibrant animation styles, entertaining heist plots, and cool, intelligent characters with clever dialogue. On the other hand, you tend to dislike movies with contrived endings... \\
    \rowcolor{gray!15} \textbf{Raw Review} \\
    \textit{[Ocean's Eleven]} : ... but the photography in "Ocean's Eleven" is, at heart, simply wonderful: tinsel colours, beguiling bright lights, tight framing ... And the heist itself is pleasingly clever. It's a charming film ...  \\
    \rowcolor{gray!15} \textbf{Reflected Preferences} \\
    \textbf{Item}: ``\textit{Ocean's Eleven}'': \\
    \textbf{Like}: Vibrant animation style, entertaining heist plot. \\
    \textbf{Dislike}: Possibility of lackluster acting, unsatisfying resolutions. \\
    \ldots \\    
    \bottomrule
    \end{tabular}%
}
\caption{An example of interactions between our user simulator and CRS (\chatgpt).}
\label{tbl:case}
\end{table}
\subsection{Qualitative Evaluation}
\subsubsection{Implementation Details} \label{appendix:qualitative_implementation}
Following \cite{liu2023g}, we employ an LLM (i.e., GPT-5) as the evaluator. 
We task the LLM with fine-grained scoring rubrics on a 1-to-5 scale, with clear criteria for each rating. 
The inputs to our qualitative evaluation process comprise generated dialogues and the general preferences unique to each user simulator. 
In assessing \proact and \coh, the LLM is instructed to carefully analyze the full dialogue history, examining how proactively the system discovers user needs while maintaining a fluent conversational tone. 
For \personal, we leverage the LLM to evaluate whether the recommender's responses, including recommendations and explanations, are consistent with the simulator's general preferences.

\section{Case Study}
Table~\ref{tbl:case} presents a dialogues generated from PEPPER, reflecting a clear alignment between our user simulator's responses and the corresponding real user preferences.
For instance, regarding the movie "Ocean's Eleven," the user mentions in their raw review an appreciation for the film's storyline, describing it as \textit{"the heist itself is pleasingly clever, it's a charming film..."}
Correspondingly, the user simulator generates reflected preferences stating: \textit{"[like] entertaining heist plot"}, which aligns with the user's original sentiments. 
Notably, the user simulator achieves this without any target item information, validating the effectiveness of our approach in representing diverse human preference and evaluating CRSs.

\clearpage
\begin{figure*}[!ht]
    \centering
    \includegraphics[width=1\textwidth]{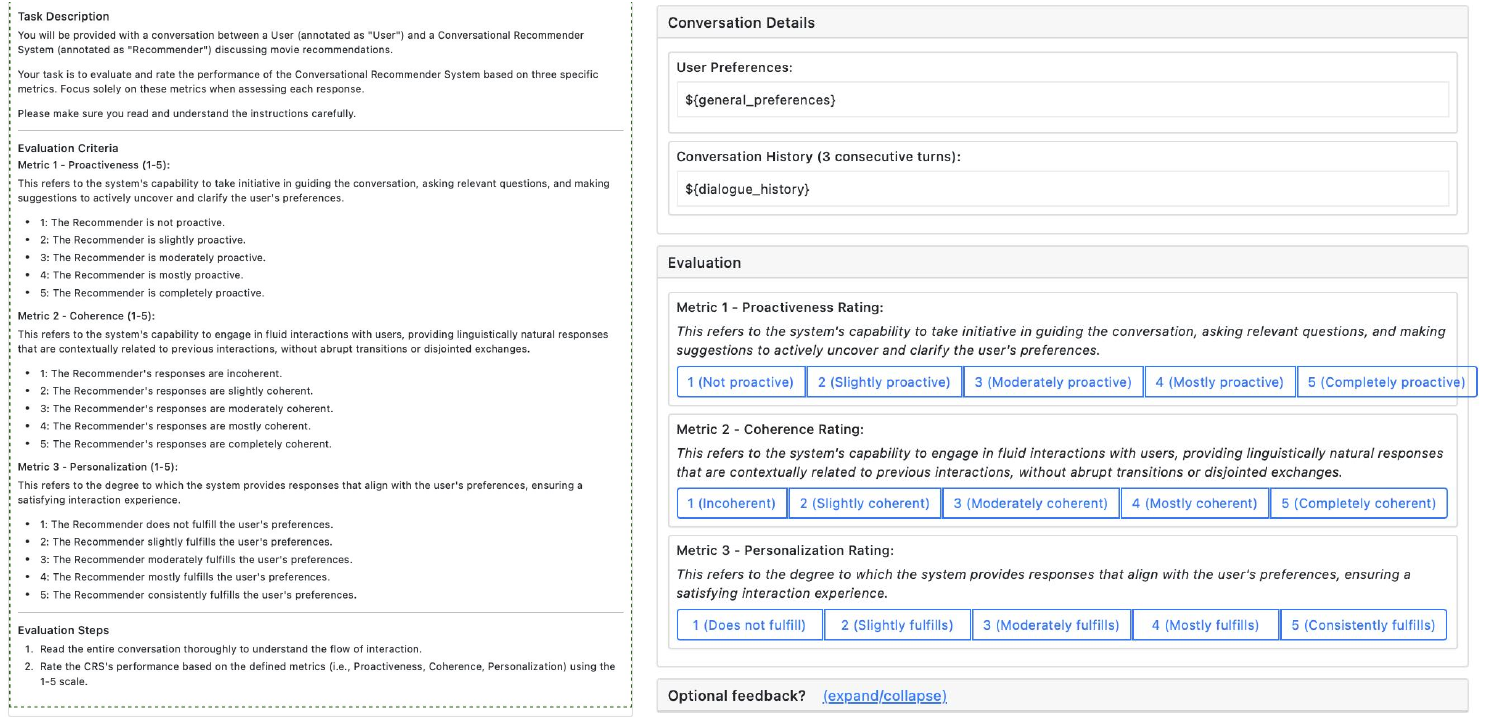}
    \caption{Human evaluation interface.
    }
    \vspace{-0.2cm}
    \label{fig:appen_human_eval}
    \vspace{-0.3cm}
\end{figure*}

\end{document}